\newcommand{\Dslash}{{D\hskip -0.23cm\slash}}

\newcommand{\tev}{\,\textrm{TeV}}
\newcommand{\gev}{\,\textrm{GeV}}

\newcommand{\tadec}{T_{\tilde a\rm -dcp}}
\newcommand{\tarad}{T_{\tilde a\rm =rad}}
\newcommand{\treh}{T_{\rm R}}

\newcommand{\td}{T_{\rm D}}

\newcommand{\axino}{{\tilde{a}}}

\newcommand{\gluino}{{\tilde{g}}}

\documentclass[12pt]{iopart}

\usepackage{graphicx,color}
\begin{document}

\title[Axinos as Dark Matter...]{
Axinos as Dark Matter Particles}

\author{Laura Covi$^1$ and Jihn E. Kim$^2$}

\address{
$^1$ Deutsches Elektronen SYnchrotron DESY, Notkestrasse 85, D-22603 Hamburg,
Germany\\
$^2$ Department of Physics and Astronomy% and Center for Theoretical Physics
, Seoul National University, Seoul 151-747, Korea
}
\ead{Laura.Covi@desy.de, jekim@ctp.snu.ac.kr}
\begin{abstract}
The identification of dark matter in our particle
physics model is still a very open question.
Here we will argue that axinos can be successful dark matter candidates in models with supersymmetry and the axion solution of the strong CP problem.
Axinos can be the lightest supersymmetric particle~(LSP), or can be heavier than the LSP.  Axinos can be produced in the right abundance by thermal scatterings and if they are the LSP also by out of equilibrium decays of the lightest superpartner of SM fields~(LSPSMs).
On the other hand heavier (not LSP) axinos can generate a part of the neutralino LSP dark matter.
Depending on the nature of the supersymmetric spectrum,  and if R-parity is strictly conserved or slightly broken, very different signals of the LSP axino scenario can arise at colliders and in astrophysics.

\end{abstract}

%Uncomment for PACS numbers title message
% Keywords required only for MST, PB, PMB, PM, JOA, JOB?
%\vspace{2pc}
%\noindent{\it Keywords}: Article preparation, IOP journals
% Uncomment for Submitted to journal title message
%\submitto{\JPA}
% Comment out if separate title page not required
\maketitle

\section{Introduction}

The nature of dark matter~(DM) as a particle is still unknown today, since its main evidence relies only on the gravitational interaction and is universal. On the other hand, from some of the present data and numerical simulations of structure formation, we do know that it must be a neutral, {\it cold}, very probably collisionless (i.e. quite weakly interacting) and very long lived particle~\cite{DMreview}. Unfortunately a particle with these characteristics is not contained in
the standard model~(SM) of particle physics: the only neutral stable and massive candidates, the electroweakly active neutrinos, are so light that they are at most hot DM and therefore only a subdominant component. DM has therefore to be part of a larger picture and of any physics beyond the SM.

The probably best motivated models of this kind rely on supersymmetry (SUSY), which is the unique extension of the Poincare' algebra and calls for a doubling of all degrees of freedom with spin difference $\Delta s=\pm \frac12$~\cite{SUSY74}. In this context it is then clear
that more particles can be suitable DM candidates, if they are the lightest one and sufficiently long-lived, in particular the very  well studied cases of the neutralino or the gravitino.

But if we invoke the Peccei-Quinn~(PQ) solution to the strong CP problem in supersymmetric models, a new multiplet has to be introduced, the axion
multiplet~\cite{Susyaxion,Frere83}. Such a multiplet must by its nature interacts with the SM particles, but the scale of its interaction is suppressed by the scale at which the PQ symmetry is broken, $F_a$. Therefore, the fermionic component of the multiplet is naturally a
very weakly interacting particle and can easily be the lightest state of  the spectrum, but it can also be heavy. We will present in this paper a summary of the axino cold DM (CDM)
scenario~\cite{CKRosz99,CKKR01,CoviRS02, CoviRAu04,Brandenburg04,Choi08,bckp08} and explore the implications of axino DM for the phenomenology of
supersymmetric models and for future indirect detection of DM.

%%%%%%%%%%%%%%%%%%%%%%%%%%%%%%%%%%%%%%%%%%%%%%%%%%%%%%%%%%%
%%%%%%%%%%%%%%%%%%%%%%%%%%%%%%%%%%%%%%%%%%%%%%%%%%%%%%%%%%
\section{Axino models and axino mass}

In discussing axino models, one should refer to the corresponding axion models. So, let us start with the axion shift symmetry and the reparametrization invariance as discussed in \cite{KimCarosi08}.

The PQ solution of the strong CP problem requires the introduction of the axion $a$, which renders the $\theta $ parameter~\footnote{Below, $\theta$ denotes the conventional $\bar\theta= \theta_{0}+{\rm Arg.Det.}m_q$.}
dynamical and allows it to relax to zero after the QCD phase transition. An axion $a$ is a pseudoscalar boson coupling to the gluon anomaly as
\begin{equation}
{\cal L}_{\theta} = \frac{\alpha_s a}{8\pi F_a}~G_{\mu\nu} \tilde G^{\mu\nu} \; ,
\end{equation}
where the dual field strength is $\tilde G^{\mu\nu}= \frac12\epsilon^{\mu\nu\rho\sigma}
G_{\rho\sigma}$ without any other interaction term in the potential $V$. Below the QCD chiral symmetry breaking scale an axion potential is developed, which arises purely from integrating out the strongly interacting fields with that anomaly term. If the original potential contains other axion dependent terms, they should be extremely small and not affect the position of the minimum, such that the axion {\it v.e.v.}
$|\theta| \sim \langle a \rangle/F_a $ should be extremely small, $<10^{-11}$.

Let us now focus on the QCD interactions containing the axionic degree of freedom. The axion effective Lagrangian below the chiral symmetry breaking scale reads
\begin{eqnarray}\hskip -1.2cm
{\cal L}_{\theta, eff}&=&
\frac12 f_{S}^2\partial^\mu \theta\partial_\mu\theta-
\frac14 G_{\mu\nu}^a G^{a\hskip 0.02cm \mu\nu}+
(\bar q_L i\Dslash q_L+\bar q_R i\Dslash q_R)+
c_1(\partial_\mu \theta)\bar q\gamma^\mu\gamma_5 q
\nonumber\\
&&-\left(\bar q_L~ m~q_R e^{ic_2\theta}+{\rm h.c.}\right)+
c_{3} \frac{\theta}{32\pi^2}G^{a}_{\mu\nu}\tilde G^{a\hskip 0.02cm\mu\nu}\
({\rm or}\ {\cal L}_{\rm det} )\label{Axionint} \\
&& +c_{\theta\gamma\gamma}\frac{\theta}{32\pi^2}F
^{i}_{\rm em,\mu\nu}\tilde F_{\rm em}^{i\hskip 0.02cm\mu\nu}
+{\cal L}_{\rm leptons,\theta} (c_1^{\ell}, c_2^{\ell})\; ,
\nonumber
\end{eqnarray}
where $\theta=a/f_S$ with the the axion decay constant $f_S$ defined up to the domain wall number ($f_S=N_{DW} F_a$) and $q$ are the SU(3)$_c$ charge carrying quark fields. The $c_1$ term is the derivative coupling with quarks respecting the PQ shift symmetry,  the $c_2$ term is related to the phase in the quark mass matrix, and the $c_3$ term is the anomalous coupling or the determinental interaction ${\cal L}_{\rm det}$. ${\cal L}_{\rm leptons,\theta}$ is the axion interaction with leptons,
which in principle can contain other constants $ c_1^{\ell}, c_2^{\ell} $. The coupling constants $c_1,c_2,$ and $c_3$ are obtained below the axion
scale $f_S$ after integrating out the heavy degrees of freedom responsible of the PQ symmetry breaking. The mass parameter $m$ is defined to be real and positive below the electroweak scale.

The Lagrangian (\ref{Axionint}) has a shift symmetry $a\to a+$ (constant), which reparametrizes the couplings $c_1,c_2,$ and $c_3$. Explicitly, changing the phases of the quark fields $q_L\to e^{i\alpha a(x)}q_L$ and $q_R\to e^{-i\alpha a(x)}q_R$, we obtain the following reparametrization, where the effective one point irreducible action
$\Gamma_{1PI}[a(x), A_\mu^a(x); c_1, c_2, c_3,  m, \Lambda_{\rm QCD}]$ changes to
\begin{eqnarray}
\Gamma_{1PI}[a(x), A_\mu^a(x); c_1 -\alpha, c_2-2\alpha, c_3+2\alpha, m, \Lambda_{\rm QCD}].
\label{Gammaone}
\end{eqnarray}
So we see immediately from this transformation for a single quark, that if it is massless, the corresponding $ c_1, c_2$ parameters disappear and we can shift away the anomaly term completely {\it with no physical effect}. This is in fact one alternative solution to the strong CP problem, and see \cite{KimCarosi08}
for a detailed discussion.

For determining the axion mass, all $c_1, c_2$ and $c_3$ terms may be relevant, but only the combination $c_2+c_3$ actually appears~\cite{KimCarosi08}.
Usually, in the field theoretic axion models, we start with $c_1=0$. In any case, note that the $c_1$ term can be reabsorbed in the $c_2$ term using integration by part and the quarks equations of motion. So, in the next sections we just start with the couplings $ c_2$ and $ c_3 $.

Usually, $F_a$ is defined by transferring all couplings of the axion to the coefficient of $G\tilde G$ and rescaling $c_3 $ to one. On the other hand, $f_S$ is defined to be the VEV of the singlet field $\sigma$ breaking the PQ symmetry. It turns out that $c_2+c_3$ is an integer, not necessarily one in the pseudoscalar
field space and it determines the number of minima in the axion periodic potential. Thus, this integer is called the domain wall number $N_{DW}$ \cite{Sikivie82}
\begin{equation}
N_{DW}=|c_2+c_3|=
{\rm Tr}\; Q_{PQ} (\psi_{\rm colored}) \ell(\psi_{\rm colored})\; ,
\label{DWnumber}
\end{equation}
where the trace is taken over all colored fermions $\psi_{\rm colored} $,
$\ell$ is the index of their SU(3)$_c$ representation and
the PQ charge $ Q_{PQ} $ is given for the left-handed chiral representations.

The scale $ F_a$ is constrained by astrophysical and cosmological bounds to lie in the narrow axion window
$ 10^{10} \mbox{GeV} < F_a < 10^{12} \mbox{GeV}$~\cite{KimCarosi08}.

Note that above the electroweak~(EW) symmetry breaking scale in principle also couplings to the EW gauge bosons and the Higgs fields may arise and we have to write then the effective Lagrangian as
\begin{eqnarray}\hskip -1.2cm
{\cal L}_{\theta, eff >EW }&=&
 \frac12 f_{S}^2\partial^\mu \theta\partial_\mu\theta-
\frac14 G_{\mu\nu}^a G^{a\hskip 0.02cm \mu\nu}+
(\bar q_L i\Dslash q_L+\bar q_R i\Dslash q_R)
\\
& & +  \partial_\mu h_I^* \partial^\mu h_I + V(h_I, \theta) + c_{1, h_I} (\partial_\mu \theta)
( h_I^* \partial^\mu h_I - \partial^\mu h_I^* h_I)
\nonumber\\
&& + c_1(\partial_\mu \theta)\bar q\gamma^\mu\gamma_5 q
- \left( Y_I^q \bar q_L~ h_I ~q_R e^{ic_2\theta}+{\rm h.c.}\right)
\nonumber\\
&& + c_{3} \frac{\theta}{32\pi^2}G^{a}_{\mu\nu}\tilde G^{a\hskip 0.02cm\mu\nu}+c_{3,Y} \frac{\theta}{32\pi^2}B
_{\rm Y,\mu\nu}\tilde B_{\rm Y}^{\mu\nu}
\nonumber\\
&&  +c_{3,EW} \frac{\theta}{32\pi^2}W
^{i}_{\rm SU(2),\mu\nu}\tilde W_{\rm SU(2)}^{i\hskip 0.02cm\mu\nu}
+{\cal L}_{\rm leptons,\theta} (c_1^{\ell}, c_2^{\ell} )\; .
\nonumber
\end{eqnarray}
Then we can define an extended shift symmetry, including also transformations of the EW charged fields; these can be changed independently to the colored degrees of freedom, so to have $ c_{3,EW} = 0 $ and leave only the anomalous coupling to the hypercharge gauge bosons.
As in the case of QCD, such a coupling could be shifted away completely if one of the leptons were massless; since the electron mass is quite small, the residual effects, contained in the $ c_{1,2}^{\ell} $ terms, is
negligible for many practical purposes.
%\vspace*{-0.2cm}

%%%%%%%%%%%%%%%%%%%%%%%%%%%%%%%%%%%%%%%%%%%%%%%%%%%%%%%%%%%%%%%%%
\subsection{Axion Models}

There are several types of $c_2$ and $c_3$ couplings which define different axion models. If $c_2=0$ and $c_3\ne 0$ due to the existence of PQ charge
carrying heavy quarks, the model is called the
Kim-Shifman-Vainshtein-Zakharov~(KSVZ) model. If the coupling $c_2$ is provided by the electroweak scale Higgs doublets, while $c_3=0$, it is the Peccei-Quinn-Weinberg-Wilczek~(PQWW) model.
If the phase $c_2$ is provided by an electroweak singlet with $c_3=0$, it is the Dine-Fischler-Srednicki-Zhitnitski~(DFSZ) model.
The model-independent axion in superstring models give instead $c_2=0$ and $c_3=1$. These values enable us to write down the axion-nucleon-nucleon couplings unambiguously for each model~\cite{KimCarosi08}.
But in general axion models may contain both $c_2$ and $c_3$ with $c_2+c_3\ne 0$ \cite{KimCarosi08} and may have the family dependencies of the variant axion~\cite{varaxion} or invisible axion~\cite{varaxionInv}.
%\vspace*{-0.2cm}

%%%%%%%%%%%%%%%%%%%%%%%%%%%%%%%%%%%%%%%%%%%%%
\subsection{Axino, SUSY breaking and axino mass}

In the case of a supersymmetric model, the axion field is the pseudoscalar part of a whole chiral multiplet $\Phi$. Note, however, that the reparametrization invariance Eq.~(\ref{Axionint}) still holds and represents a freedom in choosing the $c_1, c_2$, and $c_3$ terms. We choose here the basis where the $c_2$ term is transferred to the $c_3$ term, and hence $\Phi$ interaction is
\begin{equation}
\int d^2\vartheta {\alpha_s \over 4\sqrt{2} \pi F_a}
\Phi_a {\cal W}^\alpha {\cal W}_{\alpha}+{\rm h.c.}\; ,
\label{dim5op}
\end{equation}
where now $\Phi_a =(s+ia)/\sqrt2+\vartheta\tilde a +(F~{\rm term})$ is the chiral multiplet containing the saxion $s$ and axion $a$ and their fermionic partner the axino $\tilde a$, while ${\cal W}_\alpha$ is the vector
multiplet containing the gluino and the gluon field strength, and
${\cal W}^\alpha {\cal W}_\alpha|_{\vartheta\vartheta}= -2i\lambda^a \sigma^m\partial_m\bar\lambda^a -\frac{1}{2} G^a_{\mu\nu} G^{a\mu\nu}+\frac{i}{2} G^a_{\mu\nu} \tilde G^{a\mu\nu}+ D^2$.
Here, $\alpha_s$ is the QCD coupling constant.
An analogous interaction is present for the hypercharge gauge multiplet with the additional coupling $ c_{3,Y} $ as discussed previously.

As  long as SUSY is unbroken, the axion multiplet remains light, since it is protected by the U(1)$_{\rm PQ}$ symmetry~\cite{KimMas84,axinomass,ChunLukas95}. This symmetry implies that no supersymmetric mass parameter is allowed for the axion multiplet since, as discussed above, the axion does not have a potential $V$ (i.e. terms in the superpotential $W$ with SUSY).

Both saxion and axino masses are split from the almost vanishing axion mass if SUSY is broken, either at tree level via the {\it v.e.v.} of some scalar field in the model and mixing with the other neutralinos or via loop
diagrams involving multiplets with split masses.
The precise value of the axino mass depends on the model, specified by the SUSY breaking sector and the mediation sector to the axion supermultiplet. Most probably, the saxion mass is around the soft mass scale $M_{\rm SUSY}$. The axino mass should also be near this scale as well. But the axino mass can also be much smaller than $M_{\rm SUSY}$~\cite{Frere83,KimMas84,axinomass}
or much larger than $M_{\rm SUSY}$ \cite{ChunLukas95}.
Therefore, we take the axino mass as a free parameter.

If R-parity is not conserved, the lightest supersymmetric partner of the SM particles can decay to ordinary particles. If R-parity is conserved,  it cannot decay to ordinary SM particles, but it can decay to axino or/and gravitino if they are lighter. Thus, the axino cosmology depends crucially on the R-parity realization. Here, we consider first models with R-parity conservation and the
thermal history of the universe can be very different depending on the hierarchy between the axino mass and the mass of the LSP (of SM multiplets) $M_\chi$.
Firstly, we consider the case $m_\axino<M_\chi$ and next  $m_\axino>M_\chi$.

The cosmology of a weakly-interacting massive particle~(WIMP) and an extra-WIMP depends on several temperatures. For example, the neutralino cosmology depends on the neutralino freeze-out
temperature~\cite{LeeWein77b,Drees93} and the gravitino/axino cosmology on the reheating temperature after inflation \cite{Ellis84,BBN}. We therefore define the following temperatures relevant for the axino
cosmology:
\begin{eqnarray}
&\tadec={\rm axino\ decoupling\ temperature}\nonumber\\
&\treh={\rm reheating\ temperature\ after\ inflation} \nonumber\\
&T_{fr}={\rm neutralino\ freeze-out\ temperature}\label{Temps}\\
&\tarad={\rm axino-radiation\ equality\ temperature} \nonumber\\
&\td = {\rm radiation\ temperature\ at\ \chi\ or\ \tilde {\it a}\ decay}\nonumber
\end{eqnarray}
where note that $\td$ corresponds to a different temperature for
$m_\axino<M_\chi$ and $m_\axino>M_\chi$.

%%%%%%%%%%%%%%%%%%%%%%%%%%%%%%%%%%%%%%%%%%%%%%%%%%%%%%%%%%%
\section{Axino cosmology with $m_\axino<M_\chi$}\label{sec:axinocosmology}

Let us consider the axion supermultiplet together with the Minimal Supersymmetric Standard Model~(MSSM) fields. Then the lightest supersymmetric particle in the MSSM~(LSPSM) $\chi$ plays an important role.
The decoupling temperature of the axino supermultiplet is of the order~\cite{Raja91},
\begin{equation}
\tadec =10^{11}\gev \left(\frac{F_a}{10^{12}\gev} \right)^2\left(\frac{0.1}{\alpha_s} \right)^3.\label{Tdec}
\end{equation}
Cosmology with the saxion $s$ is a simple extension of the standard cosmology if the saxion mass is around the SUSY breaking scale \cite{Kimsaxion91} or larger \cite{Kawasaki07}, but its effect is not so dramatic as the effect of the axino. If the axion interaction ever was in thermal equilibrium, e.g. $ \treh > \tadec $, a substantial axino number density survives to the
present day and axinos have to be very light.
Axinos with mass in the eV range from this epoch have been considered as hot DM~\cite{KimMas84} or warm DM for masses in the keV range~\cite{Raja91}. In the gauge mediated SUSY breaking~(GMSB) scenario, the gravitino
is probably the LSP and the possibility of primordial axinos decaying to gravitinos has been considered as well~\cite{AxinoGravitino93}.
Here let us  focus on the CDM axino LSP scenario.

%%%%%%%%%%%%%%%%%%%%%%%%%%%%%%%%%%%%%%%%%%%%%%%%%%%%%%%%%%
\subsection{Producing axinos in the Early Universe}

We briefly review here the two main mechanisms that produce axinos in the early universe. In principle also other sources could be present like Q-balls decay~\cite{Roszkowski:2006kw}. We concentrate here on the hadronic type of axion models and define the axion supermultiplet in the basis where the $c_2$ term is zero. Then the axino does not interact directly with the MSSM multiplets apart from the gluon and hypercharge vector multiplets and does not mix substantially with the standard neutralinos or other fermions. So we neglect any interaction with the leptons or the EW gauge bosons,
that may appear in the DFSZ type of models, and that can only increase the production cross-section. It is worthwhile to recall here again that the shift between $c_2, c_2^{\ell} $ and $ c_3, c_{3,EW} $ couplings is simply a matter of definition of the axion interactions
as long as it is nearly a mass eigenstate and if we choose $ c_2, c_{3,EW}= 0 $, the other coupling $c_2^{\ell} $ is suppressed by the leptons Yukawa and therefore negligible for axino production.

\subsubsection{Thermal scatterings}

\begin{figure}
\centerline{\includegraphics[height=
 .4\textheight]{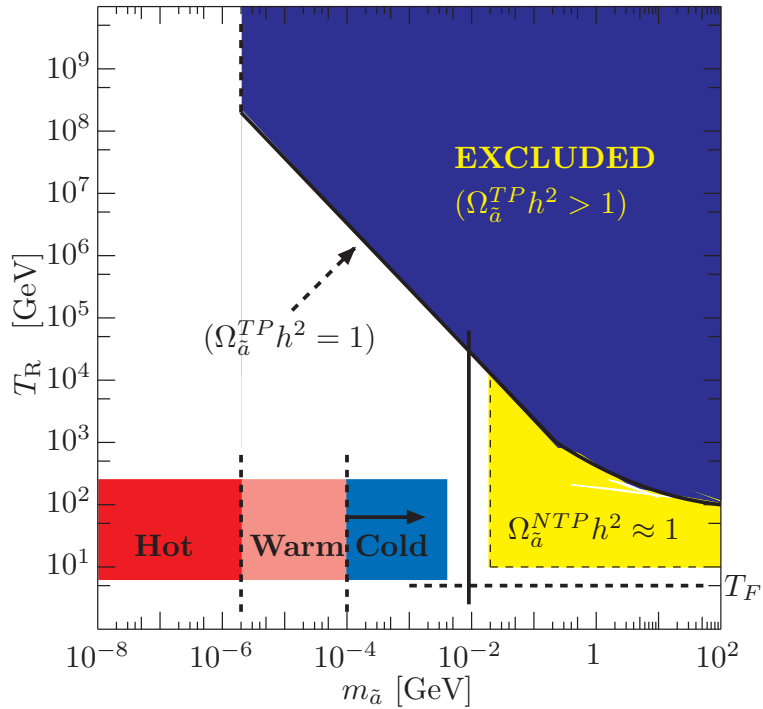}}
  \caption{Maximal reheating temperature as a function of the axino mass obtained by requiring that the axino energy density is below the present DM density~\cite{CKKR01}. The difference between solid and dashed lines is due to the inclusion of the decay term in the Boltzmann equation~(\ref{Boltzmann}). In the yellow area we expect the non-thermal production via out of equilibrium decays to be also substantial.}\label{fig:CKKR1}
\end{figure}

Any particle, even very weakly coupled, is produced in the thermal plasma by scatterings of the particles that are in thermal equilibrium. As we have seen axinos couple directly to the gluons and gluinos via the ``anomaly'' coupling in Eq. (\ref{dim5op}), i.e. in components
\begin{equation}
{\cal L}_{\tilde a g \tilde g} =
{\alpha_s \over 8\pi F_a} \bar {\tilde a} \gamma_5
\sigma^{\mu\nu} \lambda^b G^b_{\mu\nu}\; ,
\end{equation}
where $\lambda^b$ is the gluino field.
So many scatterings in the primordial plasma involving colored particles
produce axinos~\footnote{The same happens also in the case of the gravitino,
but with different vertex structure and scale \cite{gravitino}.}.
The axino number density is given by solving a Boltzmann equation of the type
\begin{eqnarray}
\!\!\!\!
{d n_{\tilde a} \over d t} + 3 H n_{\tilde a}\! &=&\!\!
\sum_{ij} \langle\sigma (i+j \rightarrow \tilde a +\dots) v_{rel}
\rangle n_i n_j
+ \sum_{i} \langle\Gamma (i \rightarrow \tilde a +\dots)\rangle n_i\, ,
\label{Boltzmann}
\end{eqnarray}
where we are neglecting back-reactions, that are suppressed by $n_{\tilde a} \ll n_i $.
At high temperature the 2-body scatterings dominate the {\it r.h.s.},
since they contain a vertex given by the dimension 5 operator in
Eq.~(\ref{dim5op}) and show a characteristic linear dependence on $T$.
So most of the axinos are produced at the highest temperature, and
the axino number density is proportional to that temperature, which we take to be $\treh$. Some of the two body scatterings are IR divergent due to the massless
gluon propagator; in the thermal bath such a divergence is screened by the presence of a thermal gluon mass $\simeq g T$. In our computation we introduced such IR cut-off by hand \cite{CKKR01}. A self-consistent procedure is instead to perform a full resummation of the hard thermal loops as done in~\cite{BS04}. In general we expect $ {\cal O}(1) $ corrections from higher orders terms in $\alpha_s $, especially at low temperature~\cite{BS04}.
There as well the decay terms start dominating and the number density is no more proportional to the reheating temperature, but depends instead on the supersymmetric spectrum, in particular the gluino and squark masses~\cite{CoviRS02}. Using the expression for the present axino energy density as
\begin{equation}
m_{\tilde a} {n_{\tilde a} (T)\over s(T)}
= 0.72\,\mbox{eV}  \left({\Omega_{\tilde a} h^2 \over 0.2 } \right)\; ,
\end{equation}
where $s(T) = 2.89\times 10^3 \left( {T \over 2.726 K} \right)
\mbox{cm}^{-3} $
is the present entropy density, we can then obtain a bound on the reheating
temperature in Fig.~\ref{fig:CKKR1}.

\subsubsection{Out of equilibrium decays}

An axino population is also generated by the LSPSM~\footnote{In passing,
we point out that the gravitino problem \cite{Ellis84} is absent if the gravitino is the NLSP, $m_{\tilde a}<m_{3/2}<m_{\chi}$, since a thermally produced gravitino would decay into an axino and an axion
which do not affect the BBN produced light elements \cite{Asaka00}.}
decay after it freezes out from the thermal bath.
The heavier superpartners cascade-decay quickly into the LSPSM (or very rarely  to the axino LSP itself as we discussed above) while still in equilibrium, but the LSPSM has a lifetime much longer than its freeze-out time: since the axino couplings are suppressed
by the PQ scale $F_a \simeq 10^{11} $ GeV, the LSPSM lifetime for $100 $ GeV mass is of the order of seconds. Then the freeze-out process is unaffected since the decay takes place only much later.

In this case, thanks to R-parity conservation, the axino energy density can be directly computed from the LSPSM would-be-relic density as
\begin{equation}
\Omega_{\tilde a}^{NT} = {m_{\tilde a}\over m_{LSPSM}} \; \Omega_{LSPSM}.
\label{omegaresc}
\end{equation}
If the mass ratio is not too small, we still have a connection with the classical WIMP mechanism in case the LSPSM is a neutralino. On the other hand in this scenario the LSPSM can be more generally any superpartner which may freeze-out with a sufficiently large number density.

A couple of problems can arise if the LSPSM decay happens too late:

\begin{itemize}
\item
Big Bang Nucleosynthesis~(BBN) can be spoiled by the energetic ``active'' particles produced in the decay along with the axino: the strong limits on the injection of energetic particles depend on the
electromagnetic/hadronic nature of the produced showers, the LSPSM number density and its decay time~\cite{BBN}. In general such limits are weak for the axino case since the LSPSM lifetime (excluding a strong mass degeneracy) is below $10^2$s, but they can affect the region of small mass for both the neutralino and stau
LSPSM~\cite{CKKR01,CoviRS02}.
This constraints disappear easily for colored LSPSM like the stop since the number density is reduced also by the Sommerfeld enhancement and the lifetime is very short as long as the decay to top is
allowed~\cite{bckp08}. Therefore a stop LSPSM is perfectly viable, but a very large stop mass and axino mass of a few TeV is needed to produce the whole DM density.

\item
Are axino from the decay cold enough to be CDM? They may be relativistic at production even if the LSPSM is not and they in general have a non-thermal spectrum. Their velocity can be estimated as:
\begin{equation}
v (T) = {p(T) \over m_{\tilde a}} \simeq
{m_{LSPSM} \over 2 m_{\tilde a}}
\left( {g_*(T ) \over g_*(\td)} \right)^{1/3} {T\over \td},
\end{equation}
where $\td $ here is the temperature of the LSPSM decay time. Axino must therefore have sufficient time to cool down before structure formation begins.
In \cite{JLM05} such constraints have been studied and the conclusion is that an axino mass of at least 1~GeV is probably needed if the whole DM population
is produced by out of equilibrium decay of a LSPSM of a 100 GeV mass.

\end{itemize}

Depending on the parameters and $\treh$, either production mechanism can give sufficient axinos to explain the present DM density. Once more information about the SUSY spectrum is available from LHC it may
be possible to determine which contribution dominates and restrict the range of $\treh $ \cite{Choi:2007rh}.
Of course another possibility is that the axino is so light to be a subdominant (warm or hot) DM component. In the last case in our scenario the axion~\cite{KimCarosi08} could be the DM.

%%%%%%%%%%%%%%%%%%%%%%%%%%%%%%%%%%%%%%%%%%%%%%%%%%%%%%%%%%%
%%%%%%%%%%%%%%%%%%%%%%%%%%%%%%%%%%%%%%%%%%%%%%%%%%%%%%%%%%%
\section{Axino cosmology with $m_\axino>M_\chi$}\label{sec:Heavyaxino}

Now, let us consider the axino mass region, $m_\axino>M_\chi$. Here, we are interested in the case where the CDM density is determined by the axino and 
in particular the axino energy density is dominating
the evolution history of the Universe. This is possible not only in the near past if the axino has not decayed yet and is DM as discussed in Sec.~\ref{sec:axinocosmology}, but also if a heavy axino decayed into the DM at an earlier epoch as will be
considered below.

Also for heavy axino, the axino density before decay can be estimated from $\tadec$ or $\treh$ as discussed above. Even in the second case, when the axion coupling never was in equilibrium, an early cold axino DM domination may have happened if the number density was sufficiently large, i.e. if $ \treh $ was larger than $\treh^{\rm min}$ defined by the equality of axino and
radiation energy density at decay:
\begin{equation}
 \frac43 m_{\tilde{a}} Y_{\tilde{a}}(\treh^{\rm min})= \td\; .
\end{equation}
So for any $ \treh > \treh^{\rm min} $ axinos dominate the evolution of the universe before they decay and produce a non-negligible amount of entropy diluting the existing number densities. We recall here that in SUSY theories  we must always consider a relatively
small reheating temperature $10^{7-8}\gev$ due to the gravitino problem~\cite{Ellis84,BBN}. The heavy axino cosmology must also satisfy this upper bound on the reheating temperature.

\begin{figure}
 \centerline{ \includegraphics[height=.5\textheight]{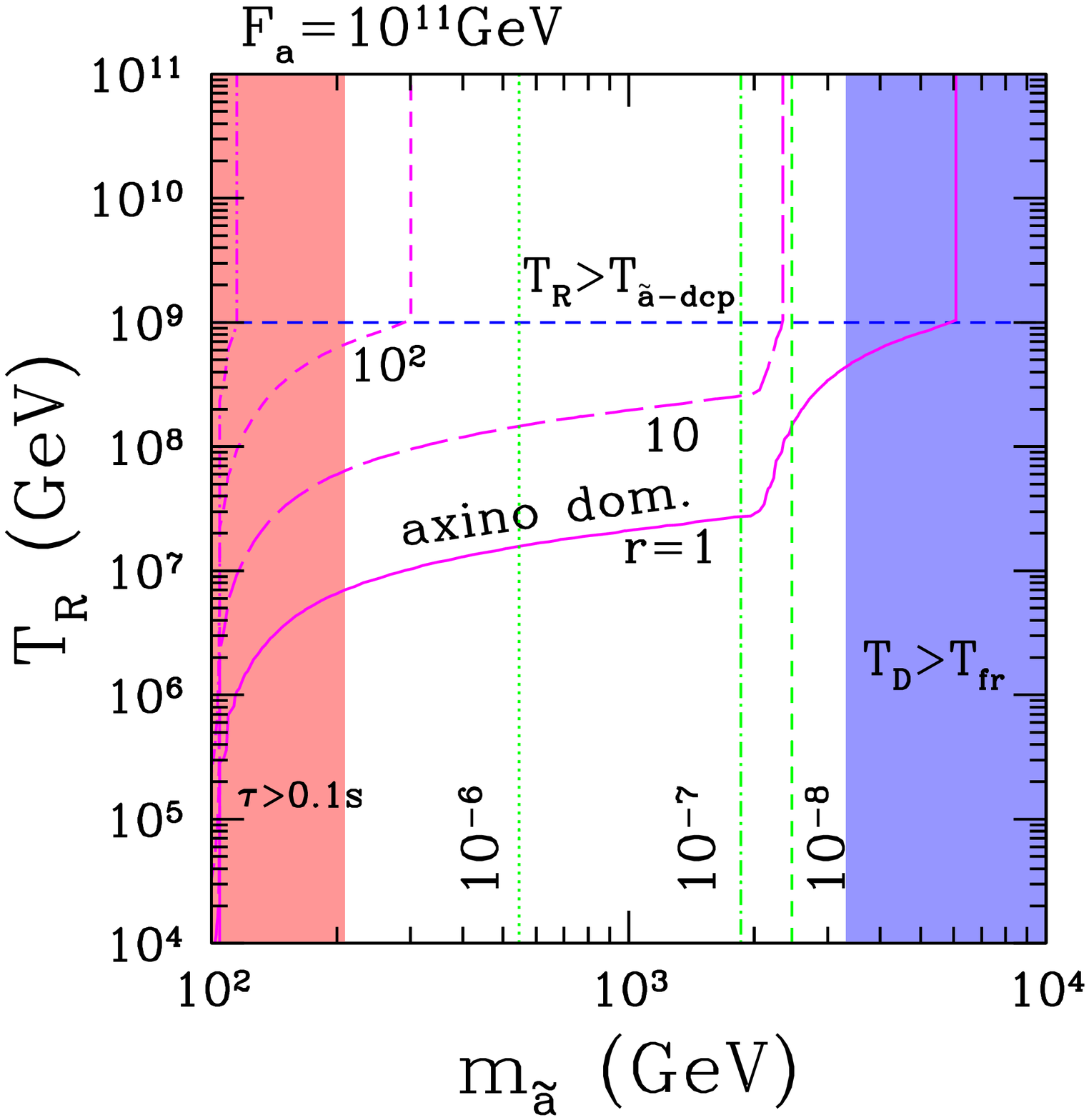}}
  \caption{
  The $\treh$ vs. $m_\axino$ plot for $m_\chi=100~\gev$ and $F_a=10^{11} \gev$.
   }\label{fig:Heavyaxino}
\end{figure}

Heavy axinos cannot be the LSP and decay to the LSP plus light SM particles. This possibility was considered briefly in studying cosmological effects of the saxion in \cite{Kawasaki07}, and a more complete cosmological
analysis has been presented in \cite{Choi08} which will be sketched here. Here the axino or the axino-decay produced neutralino is supposed to constitute the CDM fraction $\Omega_{\rm CDM}\simeq 0.23$ of the universe.

For a heavy axino decaying to a neutralino, we present a $\treh$ vs. $m_\axino$ plot for $F_a=10^{11} \gev$ in Fig.~\ref{fig:Heavyaxino}. For other parameters, we refer to~\cite{Choi08}. The region $\treh>\tadec$ is above the dashed blue line. The axino lifetime is greater than $0.1$ s in the red shaded region in the LHS and there BBN constraints may apply. The blue shaded region in the RHS is where the axino decays before the neutralino decouples ($\td > T_{fr}$). The magenta lines (horizontal) are the contours of the entropy increase due to the axino decay, $r\equiv S_f/S_0$. Above the $r=1$ line axinos dominate the universe before they decay. The green lines (vertical) show the values of $\langle \sigma_{ann} v_{rel}\rangle$, where $\sigma_{ann}$ is the
neutralino annihilation cross section, in units of ${\rm GeV}^{-2}$ which are needed to give the right amount of neutralino relic density after $T_{fr}$. In Fig.~\ref{fig:Heavyaxino}, we use neutralino
and gluino masses as $m_\chi=100\gev$ and $m_\gluino = 2 \tev$, respectively. For a larger $F_a$ and a heavier neutralino mass, the green lines move to the
right \cite{Choi08}.

%%%%%%%%%%%%%%%%%%%%%%%%%%%%%%%%%%%%%%%%%%%%%%%%%%%%%%%%%%
\section{Axino DM and R-parity breaking}

The axino is a good DM candidate even if R-parity is not exactly conserved. In fact its lifetime can be very long, thanks to the PQ scale suppression and a small R-parity breaking. Axino CDM with R-parity breaking has been
considered in \cite{HBKim02} and subsequently in the context of the Integral anomaly in \cite{HW04-Integral, CK06-Integral}. Different decay channels and lifetimes are possible for the axino CDM depending on the R-parity breaking model and the axion model. If we consider bilinear R-parity breaking of the form~\cite{bilRp}
\begin{equation}
W_{R\!\!\! \slash} = \epsilon_i \mu L_i H_u\; ,
\end{equation}
where $\mu $ is the Higgs bilinear term, and the corresponding SUSY
breaking soft term
\begin{equation}
{\cal L}_{soft R \!\!\! \slash } = B_i \mu \tilde L_i H_u ~,
\end{equation}
and restrict ourselves to hadronic axion models, the axino decay arises from the anomaly coupling with the photon vector multiplet and the neutralino-neutrino mixing generated by the sneutrino {\it v.e.v}
$\langle \tilde\nu_i \rangle $. The decay rate then reads
\begin{equation}
\Gamma_{\tilde a \rightarrow \gamma\nu_i} =
\frac{C_{a\gamma\gamma}^2 \alpha_{em}^2 m_{\tilde a}^3}{2 (4\pi)^3 F_a^2}
\xi_i^2 | U_{\tilde\gamma \tilde Z} |^2\; ,
\end{equation}
where $\xi_i = \langle \tilde\nu_i \rangle/v $
with the Higgs {\it v.e.v.} given by $ v = 174~ \mbox{GeV} $ and
\begin{equation}
U_{\tilde\gamma \tilde Z} = M_Z \sum_\alpha \frac{ S_{\tilde Z\alpha} S^*_{\tilde \gamma \alpha}}{m_{\chi_\alpha}}
\end{equation}
is the photino-Zino mixing parameter, containing the neutralino mixing matrix $ S$ and the mass eigenvalues $ m_{\tilde\chi_\alpha}$ as in the case of the
gravitino~\cite{Covi:2008jy}.
The lifetime is then given by
\begin{equation}
\tau_{\tilde a \rightarrow \gamma\nu} =
4 \times 10^{25} s
\left( \frac{\xi_i  | U_{\tilde\gamma \tilde Z} |}{10^{-10}}\right)^{-2}
\left( \frac{m_{\tilde a}}{1\;\mbox{GeV}} \right)^{-3}
\left( \frac{F_a}{10^{11} \mbox{GeV}} \right)^2\; .
\end{equation}
For larger axino masses also the decay into $Z$ bosons opens up, and quickly dominates since it is proportional to the $ U_{\tilde Z \tilde Z} $ mixing, which can be as large as one. The decay into $W$ instead does not appear since we have shifted away the SU(2) anomaly vertex.
Note that the constraints from the diffuse gamma-ray background require a very small value for the parameters $\xi_i $ and therefore a very small R-parity breaking, smaller than in the case of the gravitino
DM~\cite{gravitinoRpv}. In fact the EGRET diffuse flux already limits the lifetime of a DM particle decaying into a single gamma line to be larger than $ \tau > 7\times 10^{26} \mbox{s} $~\cite{bbci}, between 0.1 -10 GeV, with very weak dependence on the mass scale,
while for lower masses bounds of the order of $ 10^{27} \mbox{s} $ are obtained from X-ray data~\cite{decayingDM}.

If the axion model is instead of the DFSZ type, direct couplings with the leptons and Higgses arise from the $\mu $ or $\epsilon_i \mu $ terms. Then a direct mixing between axino-neutrino and axino-higgsino appears and other channels open up, in particular the 3-body decay into neutrino and $ \ell^+ \ell^- $ via intermediate $Z$, which could contribute to the electron flux~\footnote{Note that also the diagram with off-shell
photon splitting into an electron position pair
generate this channel, but it is then subleading in comparison to the $\gamma \nu $ two-body channel due to $\alpha_{em} $ and phase-space suppression.}.
The mixing of the axino with leptons may arise actually also in KSVZ models at the one loop level~\cite{CoviRAu04}. Taking the mixing between the axino and Higgs multiplet typically as $ v/F_a $ and the Higgsino mixing to neutrino from the sneutrino {\it v.e.v.} as above, we obtain for this channel
\begin{eqnarray}
\Gamma_{\tilde a \rightarrow e^+ e^- \nu_i} &= &
\frac{G_F^2 m_{\tilde a}^5 }{192\pi^3}
\frac{|U_{\tilde H \tilde Z}|^2 \xi_i^2 v^2 }{F_a^2}
\\
& \sim & (10^{26} \mbox{s} )^{-1}
\left(\frac{m_{\tilde a}}{10\;\mbox{GeV}} \right)^5
\left(\frac{F_a}{10^{11}\;\mbox{GeV}} \right)^{-2}
\left(\frac{|U_{\tilde H \tilde Z}| \xi_i}{10^{-12}} \right)^2\; ,
\nonumber
\end{eqnarray}
where $G_F $ is the Fermi constant and $ U_{\tilde H \tilde Z} $ the mixing between Higgsino and Zino neutralino defined in an analogous way as $ U_{\tilde \gamma \tilde Z} $. The branching ratios in the different leptons and quarks are determined by the axino mass and the $Z$ couplings. In general the same decay can also arise via the R-parity breaking leptonic trilinear coupling $ \lambda L L E^c $, and then the decay rate is given by
\begin{eqnarray}\hskip -1.2cm
\Gamma_{\tilde a \rightarrow e^+ e^- \nu} &=&
\frac{\zeta_e^2 |\lambda |^2 m_{\tilde a}^5}{24 (8\pi)^2 F_a^2
m_{\tilde e_R}^2}\\
&=&
( 10^{25} \mbox{s} )^{-1}
\left(\frac{m_{\tilde a}}{10\;\mbox{GeV}} \right)^5
\left(\frac{m_{\tilde e_R}}{100\;\mbox{GeV}} \right)^{-2}
\left(\frac{F_a}{10^{11}\;\mbox{GeV}} \right)^{-2}
\left(\frac{\zeta_e |\lambda |}{10^{-12}} \right)^2\; ,
\nonumber
\end{eqnarray}
in the limit $ m_{\tilde e_R} \gg m_{\tilde a} $, where we have taken
$\zeta_e m_{\tilde e_R}/F_a $ as the effective coupling of the axino to the electron multiplet. Which of the two leptonic diagrams dominate depends on the R-parity breaking and axion model parameters.

Recently, the PAMELA satellite experiment reported a significant positron excess~\cite{PAMELAe}, but no antiproton excess~\cite{PAMELAp}. If this result is confirmed by another independent experiment, a vast
unknown realm of the CDM cosmology will open up, because it is generally very contrived to build SUSY models producing excess positrons, but no excess antiprotons.
It is clear from the above discussion though, that the decaying axino could be a possibility, since in that case the radiative or leptonic decay channels may be preferred over the hadronic ones~\footnote{The leptophilic coupling for the axino was also observed
in flipped SU(5) models \cite{varaxionInv,Huh08}.}.
In fact in the bilinear R-parity violating case, the two body decays into $ W  e^+ $ is either not open (for the hadronic axion models) or may be suppressed and therefore the antiproton flux from $W$ fragmentations
disappears leaving only the $Z$ contribution.
The direct channel into $ e^+ e^- \nu $ may be dominant for models where the sneutrino {\it v.e.v.}, i.e. the bilinear R-parity breaking, is suppressed and the trilinear R-parity violating couplings give the
dominant decay.

The model-independent case of a fermion decaying
into $ e^+ e^- \nu $  has been recently studied in the context of the PAMELA anomaly in \cite{it09}, and there it was shown that such decaying DM particle may be a good fit to the data for lifetime around $10^{26} $ s and mass above 300 GeV. The axino CDM could be a realization of this scenario with the appropriate choice of parameters.
In that case though the axino has to be pretty heavy and therefore the reheating temperature very low. Even larger masses are probably needed for trying to accommodate the ATIC anomaly as well~\cite{atic08}. Note though that the PAMELA excess could be also due to astrophysical sources
like pulsars~\cite{pulsar} and then the PAMELA data give only a bound on the axino lifetime and R-parity breaking parameters. In that case also the possibility of a heavy axino $m_\axino>M_\chi$ with a neutralino DM, which cannot fit very easily with the PAMELA anomaly, is not ruled out either.

%%%%%%%%%%%%%%%%%%%%%%%%%%%%%%%%%%%%%%%%%%%%%%%%%%%%%%%%%%
\section{The LSPSM and colliders}

The signal of axino DM at colliders depends strongly on the nature of the LSPSM, which in turn depends on the SUSY breaking mechanism. In the constrained MSSM, where all the SUSY breaking parameters are derived by two common mass parameters, $m_0, m_{1/2} $, and a common
trilinear coupling $ A$ at the unification scale, the value of $\tan\beta $ and the sign of $\mu $, the only allowed LSPSMs are the lightest neutralino and the stau. In more general SUSY breaking models, of course
other LSPSMs are allowed, in particular in the case of non universal Higgs scalar masses, the stop or the sneutrino.

If the neutralino or the sneutrino are the LSPSM, it will be difficult to disentangle the two and prove that they are not DM. In both cases it would be necessary to measure their mass and couplings and realize that
those parameters either give a too large DM energy density or are already excluded by direct DM searches.
Then we would have good reasons to imply that the neutralino or sneutrino must be unstable on cosmological timescales, but it will be very difficult to determine what they are decaying into and if that includes the axino. Other, more indirect, collider signatures may arise in models with axino DM and $SO(10) $ Yukawa unification~\cite{Baer:2008eq}.

If the stau (or another charged sparticle) is the LSPSM instead, we will have the striking signal of an apparently stable charged heavy particle in the detector. In that case it will be clear that the LSP must be a
very weakly interacting particle or that R-parity is violated, but we will need to measure and study the LSPSM decay to distinguish the two possibilities and identify if there is a DM candidate and which kind of particle it is. Unfortunately the astrophysical constraints on the R-parity violation scenarios discussed in the previous section seem to point to a quite long LSPSM lifetime, if the axinos are DM, and the decay would mostly happen outside the detector.

%%%%%%%%%%%%%%%%%%%%%%%%%%%%%%%%%%%%%%%%%%%%%%%%%%%%%%%%%%%%
\subsection{How to distinguish the LSP from LSPSM decay ?}

The LSPSM decay can give information on the scenario and on the nature of the LSP, even if the LSP is not detected. In fact, the decay time and the branching ratios are model dependent and vary substantially e.g. between R-parity conserving and R-parity violating scenarios. In the first case, we expect that the dominant decay is the two-body channel into the LSPSM partner and the axino, while the next open channel the subleading radiative decay with an additional photon in the final state. If instead R-parity is violated, the LSPSM decays completely into SM particles with no missing energy apart for the light neutrinos. So for the case of a stau LSPSM, we have
\begin{eqnarray}
\tilde \tau \rightarrow \tau\; \tilde a, \tau\; \tilde a\; \gamma
\quad\quad\quad\quad\quad\quad\quad\quad  \mbox{R-parity conserved;} \\
\tilde \tau \rightarrow \tau\;  \nu_\mu, \mu\; \nu_{\tau},
b\; t^c (b\; b^c W^- ) \quad\quad\,
\mbox{R-parity violated;}
\end{eqnarray}
therefore the R-parity violation case should clearly be visible via the large lepton number breaking, since e.g. the $\tau $ and $\mu $ final states arise from the same trilinear coupling and as well from the
hadronic channel~\cite{RpVchannel}.

Moreover the angular distribution of the radiative decays into photon, a SM particle and missing energy, contains in general information on the spin of the LSP and the interaction vertex structure. This quantity can indeed play a key role in particular in order to distinguish
between axino or gravitino LSP, that can give rise to similar NLSP lifetimes and similar ``visible" decay channels~\cite{Buchmuller:2004rq, Brandenburg04}.
In that particular case we will need to measure the branching ratio and the angular dependence of the radiative decay in order to reach a definitive
identification~\cite{Brandenburg04}.

%%%%%%%%%%%%%%%%%%%%%%%%%%%%%%%%%%%%%%%%%%%%%%%%%%%%%%%%%%%
\section{Conclusions}

We have discussed here different cosmological scenarios where the axinos play an important role in the DM question. If they are light, with masses in the MeV-GeV range, they can be the CDM if the reheating~temperature is low and they are the LSP. In that case they can remain DM even if R-parity is broken, but the breaking has to be very suppressed. If instead axinos are heavy and not the LSP, they can still produce the necessary neutralino LSP
abundance in their decay and dilute dangerous relics.

In general, the presence of an axino LSP and DM relaxes many of the bounds on the supersymmetric parameters, since the right number density of axinos can be obtained in a wider region of parameter space. Moreover the possibility of different LSPSMs and therefore very different collider signature arises. We expect LHC will soon clarify the situation. In the case of R-parity violation also astrophysical signatures could arise, but they are unfortunately strongly dependent on the axion model realization.

\vspace*{-0.2cm}

\section*{Acknowledgments}

It is a pleasure to thank C. Berger, A. Brandenburg, K.-Y. Choi, K. Hamaguchi, H.B.~Kim, S. Kraml, B. Kyae, F. Palorini, R. Ruiz de Austri, M. Small, F.D. Steffen and in particular L. Roszkowski for several years of fruitful and exciting collaboration.

LC acknowledges the support of the ``Impuls- und Vernetzungsfond" of the Helmholtz Association under the contract number VH-NG-006 and of the European Network of Theoretical Astroparticle Physics ILIAS/N6
under contract number RII3-CT-2004-506222.
JEK acknowledges the Korea Research Foundation under Grant No. KRF-2005-084-C00001 of Ministry of Education, Science and Technology~(MEST) of Republic of Korea for financial support.

\end{document}